\documentclass[twocolumn,english,pra,eqsecnum,unsortedaddress]{revtex4-1}
\usepackage[T1]{fontenc}
\usepackage[latin9]{inputenc}
\usepackage{amsmath}
\usepackage{amssymb}
\usepackage{graphicx}
\usepackage{esint}

\providecommand{\tabularnewline}{\\}

\usepackage{babel}
\begin{document}

\title{Planar quantum squeezing and atom interferometry}

\author{Q. Y. He}

\affiliation{ARC Centre of Excellence for Quantum-Atom Optics, Centre for Atom
Optics and Ultrafast Spectroscopy, Swinburne University of Technology,
Melbourne 3122, Australia}

\author{Shi-Guo Peng}

\affiliation{Department of Physics, Tsinghua University, Beijing 100084, China}

\author{P. D. Drummond}

\email{pdrummond@swin.edu.au}

\affiliation{ARC Centre of Excellence for Quantum-Atom Optics, Centre for Atom
Optics and Ultrafast Spectroscopy, Swinburne University of Technology,
Melbourne 3122, Australia}

\author{M. D. Reid}

\email{mdreid@swin.edu.au}

\affiliation{ARC Centre of Excellence for Quantum-Atom Optics, Centre for Atom
Optics and Ultrafast Spectroscopy, Swinburne University of Technology,
Melbourne 3122, Australia}
\begin{abstract}
We obtain a lower bound on the sum of two orthogonal spin component
variances in a plane. This gives a novel planar uncertainty relation
which holds even when the Heisenberg relation is not useful. We investigate
the asymptotic, large $J$ limit, and derive the properties of the
planar quantum squeezed states that saturate this uncertainty relation.
These states extend the concept of spin squeezing to any two conjugate
spin directions. We show that planar quantum squeezing can be achieved
experimentally as the ground state of a Bose-Einstein condensate in
two coupled potential wells with a critical attractive interaction.
These states reduce interferometric phase noise at all phase angles
simultaneously. This is useful for one-shot interferometric phase-measurements
where the measured phase is completely unknown. Our results can also
be used to derive entanglement criteria for multiple spins $J$ at
separated sites, with applications in quantum information.
\end{abstract}


\maketitle

\section{Introduction}

Heisenberg's famous uncertainty relation for angular momentum takes
the well-known product form $\Delta J_{X}\Delta J_{Y}\geq|\langle\hat{J}_{Z}\rangle|/2$.
If $\langle\hat{J}_{Z}\rangle=0$, as in a singlet state, the Heisenberg
relation becomes trivial: it only constrains the spin variances to
positive values. Nevertheless, there are still fundamental limits
to these uncertainties, which are directly related to quantum limits
on interferometric phase measurements \cite{spinsqkiti,spinsqwine,oberthaler,oberthaler2010,atom-chip},
together with problems like macroscopic entanglement \cite{vedral,hoftoth,spinsq,toth2009},
Bohm's Einstein-Podolsky-Rosen (EPR) paradox \cite{eprbohmparadox}
and steering \cite{hw-1-1,wiseman2007,cavalsteerepr,naturephysteer}. 

Reduction of quantum noise in one spin component - or single-axis
squeezing - is a valuable tool for enhancing the sensitivity of interferometers
and atomic clocks \cite{spinsqkiti,spinsqwine}. It has been recently
implemented for ultra-cold atomic Bose-Einstein condensate (BEC) interferometers
\cite{oberthaler,oberthaler2010,atom-chip}. This type of quantum
noise reduction reduces the measurement noise near some predetermined
phase. However, if the phase is completely unknown prior to measurement,
then it is not known which phase quadrature should be in a squeezed
state. 

In this paper we demonstrate that spin operators permit another type
of quantum squeezing that we call planar quantum squeezing (PQS),
which simultaneously reduces the quantum noise of two orthogonal spin
projections below the standard quantum limit of $J/2$, while increasing
the noise in a third dimension. This allows the prospect of improved
phase measurements at any phase-angle. PQS states that reduce fluctuations
everywhere in a plane have potential utility in `one-shot' phase measurements,
where iterative or repeated measurement strategies cannot be utilized. 

Planar quantum squeezing is related to a planar uncertainty relation,
true for any quantum state. This has the form of a lower bound on
the planar spin variance sum \cite{finkel,hoftoth}, 
\begin{equation}
\Delta^{2}\mathbf{J}_{\parallel}\geq C_{J}\,.\label{eq:twospinsumcj}
\end{equation}
Here $\mathbf{J}_{\parallel}$ is the spin projection parallel to
a plane, so in the $X-Y$ plane, $\Delta^{2}\mathbf{J}_{\parallel}\equiv\Delta^{2}J_{X}+\Delta^{2}J_{Y}$,
and $C_{J}$ is the minimum of the uncertainty sum for quantum states
with fixed spin $J$. While values of $C_{J}$ for $J=1/2$ \cite{finkel}
and $J=1$ \cite{hoftoth} were known previously, we calculate $C_{J}$
for arbitrary spin quantum number $J$. We show that the $C_{J}$
uncertainty principle has a fractional exponent behavior, with $C_{J}\sim J^{2/3}\,,$
and that states saturating this uncertainty condition have variances
with the same fractional power law exponents in two orthogonal directions.
These also have a mean spin vector in the direction of greatest variance
reduction. The variance perpendicular to the squeezing plane is increased,
with $\Delta^{2}J_{\perp}\sim J^{4/3}$. The overall three-dimensional
variance has an ellipsoidal shape, graphed in Fig. 1. We show that
the ground state of a two-mode Bose-Einstein condensate (BEC) with
attractive interactions gives the maximum possible PQS, making this
an important candidate for interferometric measurements.

\begin{figure}
\begin{centering}
\includegraphics[width=0.5\columnwidth]{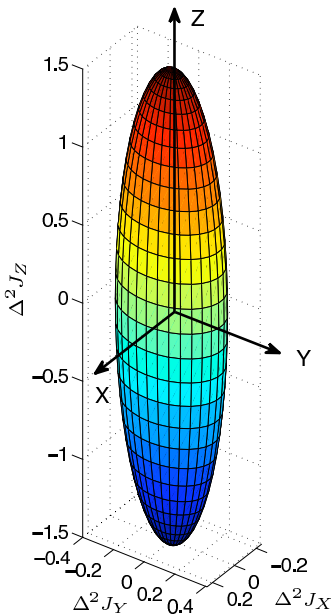}
\par\end{centering}

\caption{(Color online) Three-dimensional plot of uncertainties for planar
quantum squeezing. The figure corresponds to $100$ atoms ( $J=50$)
in the ground-state of a double-well Bose-Einstein condensate with
critical attractive interaction. Spin variances are reduced in both
axes parallel to the $X-Y$ plane, with a maximum variance reduction
in the mean spin direction $X$. The variance must increase perpendicular
to the squeezing plane, along the $Z$ axis. }
\end{figure}

As well as allowing improved interferometric phase measurements, planar
uncertainty relations are useful for the detection of non-classical
behavior in mesoscopic systems with large total spin $J$. The general
form of the corresponding two-spin Local Uncertainty Relation (LUR)
criterion to detect entanglement among $N$ sites of spin $J$ particles
is 
\begin{equation}
\Delta^{2}\mathbf{J}_{\parallel}^{total}<NC_{J}\,,\label{eq:twospinentcj}
\end{equation}
where $J^{total}$ represents the collective spin operator, and the
particular case of $N=2$ has been considered by Hofmann and Takeuchi
\cite{hoftoth} as a criterion for entanglement between two sites.
Larger $N$ values signify multipartite entanglement. Experiments
involving measurements in two spin directions \cite{expphotonweinfurt}
have employed a similar inequality to detect genuine multipartite
entanglement in four qubit states \cite{tothjosaBdickefour}. 

We extend this microscopic analysis to mesoscopic spins of \emph{arbitrary}
$J$, and hence show that equation (\ref{eq:twospinentcj}) is a multipartite
entanglement criterion for $N$ sites. This entanglement signature
is valid in the mesoscopic limit of large $J$, and applies regardless
of the third component, which may not be readily measurable.

\section{Planar Uncertainty Relations}

Our first task is to find the lower bound $C_{J}$ of the planar uncertainty
relation equation (\ref{eq:twospinsumcj}). We consider states of
fixed spin dimensionality $J$. The most general pure quantum state
of this type has dimension $d=2J+1$. Expressed in the $J_{Z}$ basis,
this state is:
\begin{equation}
|\psi\rangle=\frac{1}{\sqrt{n}}\sum_{m=-J}^{J}R_{m}e^{-i\phi_{m}}|J,m\rangle\label{eq:purestate}
\end{equation}
 Here $R_{m},\phi_{m}\ (m=-J,...,J)$ are real numbers indicating
amplitude and phase respectively, and the normalization coefficient
is $n=\sum_{m=-J}^{J}R_{m}^{2}$.

\subsection{Uncertainty Minimization}

To obtain a lower bound on the variance sum we minimize the uncertainty
over all possible expansion coefficients, using:
\begin{equation}
\Delta^{2}\mathbf{J}_{\parallel}\equiv\left\langle \left|\hat{\mathbf{J}}_{\parallel}\right|^{2}\right\rangle -\left|\left\langle \hat{\mathbf{J}}_{\parallel}\right\rangle \right|^{2}.\label{eq:spinvariance}
\end{equation}
Due to spherical symmetry, it is enough to consider the uncertainty
relation in the $X-Y$ plane. We can choose axes in the $X-Y$ plane
so that $\langle\hat{J}_{Y}\rangle=0$, with no loss of generality.
We calculate the expectation values in the $Z$ basis. We find that
the squared projections are 
\begin{eqnarray}
\left\langle \left|\hat{\mathbf{J}}_{\parallel}\right|^{2}\right\rangle  & = & -\frac{1}{4}+\frac{1}{n}\sum_{m=-J}^{J}R_{m}^{2}\left[\tilde{J}^{2}-m^{2}\right]\,,\label{eq:meansquarevalue}
\end{eqnarray}
where we have defined $\tilde{J}\equiv J+1/2$. On maximizing the
magnitude with respect to the phase variable, and introducing $M\equiv m+1/2$,
$M_{\pm}=M\pm1/2$, we find that the mean projections satisfy:
\begin{equation}
\langle\hat{J}_{X}\rangle=\pm\frac{1}{n}\sum_{M=-\tilde{J}}^{M=\tilde{J}}R_{M_{+}}R_{M_{-}}\sqrt{\tilde{J}^{2}-M^{2}}\,.\label{eq:meanvalue}
\end{equation}

Using these equations, we can numerically obtained the value of $C_{J}$
for any spin $J$, by using nonlinear optimization techniques (we
used the quasi-Newton method implemented via the Mathematica 8.0 \emph{FindMinimum}
function\cite{mathematica}) to search for the minimum value of equation
(\ref{eq:spinvariance}) given any possible coefficients. 

\begin{figure}
\begin{centering}
\includegraphics[width=0.9\columnwidth]{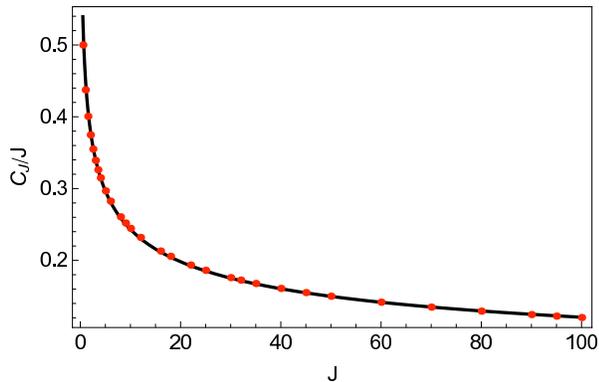} 
\par\end{centering}

\caption{(Color online) Graph of planar uncertainty relation bound.\textbf{
}The figure plots the lower bound to the planar uncertainty relation,
normalized by $J$, the standard quantum limit. Note that $C_{J}/J$
shows a decrease with increasing $J$. Discrete points are calculated
numerically, while the solid line is the analytic approximation $C_{J}^{(a)}$.\label{fig:cj-1}}
\end{figure}

The results for selected values of $J$ are tabulated in Table \ref{tab:Lower-bound-cj},
and $C_{J}/J$ is graphed in Fig. \ref{fig:cj-1} by dots.

\begin{table}
\begin{centering}
\begin{tabular}{|c|c|}
\hline 
$J$  & $C_{J}$ \tabularnewline
\hline 
\hline 
$\frac{1}{2}$  & $1/4$ \tabularnewline
\hline 
$1$  & $7/16$ \tabularnewline
\hline 
$\frac{3}{2}$  & $0.6009$ \tabularnewline
\hline 
$2$  & $0.7496$ \tabularnewline
\hline 
$\frac{5}{2}$  & $0.8877$ \tabularnewline
\hline 
$3$  & $1.018$ \tabularnewline
\hline 
$\frac{7}{2}$  & $1.142$ \tabularnewline
\hline 
$4$  & $1.260$ \tabularnewline
\hline 
$5$  & $1.484$ \tabularnewline
\hline 
$6$  & $1.695$ \tabularnewline
\hline 
$7$  & $1.894$ \tabularnewline
\hline 
$10$ & $2.445$\tabularnewline
\hline 
$20$ & $3.984$\tabularnewline
\hline 
$50$ & $7.503$\tabularnewline
\hline 
\end{tabular}
\par\end{centering}

\caption{Lower bound $C_{J}$ of the planar uncertainty equation. Numerical
results for equation (\ref{eq:twospinsumcj}) are tabulated as a function
of spin $J$. \label{tab:Lower-bound-cj}}
\end{table}

\subsection{Asymptotic $C_{J}$ values:}

We wish to obtain the asymptotic, large $J$ limit of the planar uncertainty
principle using an integral approximation. The mean spin projections
for any quantum state are:
\begin{eqnarray}
\langle\hat{J}_{X}\rangle & = & \frac{1}{n}\sum_{m=-J}^{J}\{\sqrt{(J-m)(J+m+1)}\,\nonumber \\
 &  & \;\;\;\;\times\;\; cos(\phi_{m}-\phi_{m+1})R_{m}R_{m+1}\}\,,\nonumber \\
\langle\hat{J}_{Y}\rangle & = & \frac{1}{n}\sum_{m=-J}^{J}\{\sqrt{(J-m)(J+m+1)}\,\nonumber \\
 &  & \;\;\;\;\times\;\; sin(\phi_{m}-\phi_{m+1})R_{m}R_{m+1}\}\,,\nonumber \\
\langle\hat{J}_{Z}\rangle & = & 0\,.\label{eq:meanspins}
\end{eqnarray}
For $J\rightarrow\infty$ we replace the summation over $2J+1$ discrete
values of $m$ by an integral over a continuous range of $\left(-1,1\right)$.
We define $\tilde{J}=J+1/2$, and introduce scaled variables of $x=m/\tilde{J}$
and $r(x)=R_{m}\sqrt{\tilde{J}}$, so that the normalization integral
becomes:
\begin{eqnarray}
n=\sum_{m=-J}^{J}R_{m}^{2} & \approx & \int_{-1}^{1}r^{2}(x)dx\,.
\end{eqnarray}
For the mean squared spin vector, taking equation (\ref{eq:meansquarevalue})
in the limit of $J\rightarrow\infty$ one obtains
\begin{equation}
\left\langle \left|\hat{\mathbf{J}}_{\parallel}\right|^{2}\right\rangle \approx-\frac{1}{4}+\frac{1}{n}\int_{-1}^{1}r^{2}(x)J^{2}(x)dx\,,
\end{equation}
where $J(x)\equiv\tilde{J}\sqrt{1-x^{2}}$. 

For the mean value term $\left\langle \hat{\mathbf{J}}_{\parallel}\right\rangle $
in equation (\ref{eq:meanvalue}), we also replace summations by integrals
in the limit of $J\rightarrow\infty$ and define a slightly modified
variable $x=M/\tilde{J}$. Next, using a Taylor expansion, we obtain:
\begin{equation}
r(x\pm\frac{1}{2\tilde{J}})=r(x)\pm\frac{1}{2\tilde{J}}\frac{dr(x)}{dx}+\frac{1}{8\tilde{J}^{2}}\frac{d^{2}r(x)}{dx^{2}}+O(\frac{1}{\tilde{J}^{3}})\,.
\end{equation}

After integrating by parts, and defining $r'(x)=dr(x)/dx$, this can
be expressed as a variational calculus problem. We minimize $V\left[r\right]\equiv\Delta^{2}\mathbf{J}_{\parallel}$
as a function of $r(x)$, where
\begin{eqnarray}
V\left[r\right] & \approx & -\frac{1}{4}+\int_{-1}^{1}r^{2}(x)\frac{J^{2}(x)}{n}dx\\
 & - & \left[\int_{-1}^{1}\left(r^{2}(x)-\frac{\left(r'(x)\right)^{2}}{2\tilde{J}^{2}}+\frac{xr(x)r'(x)}{4J^{2}(x)}\right)\frac{J(x)dx}{n}\right]^{2}\,.\nonumber 
\end{eqnarray}
Due to the symmetry of the integrand around $x=0$, the minimum function
$r(x)$ must have a reflection symmetry with a maximum at $x=0$.
In the large $J$ limit we can assume a relatively narrow Gaussian
solution of variance $\sigma\ll1$, where $\sigma\sim J^{-\nu}$,
$\nu$ is still undetermined, and: 
\begin{eqnarray}
r(x) & = & (\frac{1}{2\pi\sigma})^{1/4}e^{-x^{2}/4\sigma}\,.
\end{eqnarray}
With this choice, we can extend the integration limits to $x=\pm\infty$
to leading order in $1/J$. This means that $n=1$, and on expanding
the integrand one finds that:
\begin{eqnarray}
V\left[r\right] & \approx & \tilde{J}^{2}\int_{-\infty}^{\infty}r^{2}(x)\left[1-x^{2}\right]dx\\
 & - & \left[\tilde{J}\int_{-\infty}^{\infty}r^{2}(x)\left(1-\frac{x^{2}}{8\sigma^{2}\tilde{J}^{2}}\right)\sqrt{1-x^{2}}dx\right]^{2}\,.\nonumber 
\end{eqnarray}
Next, we expand the variational integral, retaining leading terms
only. We use the results that:
\begin{eqnarray}
\int_{-\infty}^{\infty}r^{2}(x)x^{2}dx & = & \sigma\nonumber \\
\int_{-\infty}^{\infty}r^{2}(x)x^{4}dx & = & 3\sigma^{2}\nonumber \\
\int_{-\infty}^{\infty}r^{2}(x)\sqrt{1-x^{2}}dx & = & 1-\frac{\sigma}{2}-\frac{3\sigma^{2}}{8}+...,
\end{eqnarray}
giving a corresponding asymptotic estimate for $C_{j}$ of 
\begin{equation}
V\left[r\right]\approx\frac{1}{4\sigma}\left[1+2\sigma^{3}J^{2}\right]...
\end{equation}
Applying variational calculus so that $dV/d\sigma=0$, and solving
in the limit of large $J$, we find that $\sigma=\left(2J\right)^{-2/3}$.
This means that: 
\begin{eqnarray}
\lim_{J\rightarrow\infty}C_{J} & \approx & 3(2J)^{2/3}/8=0.595275J^{2/3}\,.
\end{eqnarray}

Using this asymptotic form, and numerically fitting the tabulated
results with a series in $J^{1/3}$, we obtain the following analytic
approximation to $C_{J}$: 
\begin{equation}
C_{J}^{(a)}\simeq0.595275J^{2/3}-0.1663J^{1/3}+0.0267\,.\label{eq:Cj}
\end{equation}
This is given in Fig. \ref{fig:cj-1} by the solid curve, in good
agreement with our numerical results: within $1\%$ for $J\geq5$.

\section{Characterization of the minimum variance state }

We now wish to characterize the properties of the state that minimizes
a planar variance sum. The main difference can be seen if one considers
that not all the states that minimize the Heisenberg product have
planar squeezing. For example, the state with $\langle\hat{J}_{Z}\rangle=-J$
is an eigenstate of the spin projection in the $Z$ direction. This
gives a minimum variance (zero) in one direction, but cannot minimize
the variance sum. It is not a planar squeezed state, since the orthogonal
variances are not reduced below the standard quantum limit.

\subsection{Optimum planar squeezed state:}

We wish to analyze the detailed asymptotic properties of the planar
quantum squeezed state that saturates the uncertainty principle. We
consider states of fixed spin dimensionality $J$, and calculate the
mean value of $\langle\hat{\mathbf{J}}\rangle$ in the $J_{Z}$ basis
as in equation (\ref{eq:meanspins}). As previously, we choose axes
with squeezing in the $X-Y$ plane and $\langle\hat{J}_{Y}\rangle=0$.
By selecting the phase as $\phi_{m}=0$, we get a minimum planar variance,
and a non-vanishing mean value in $\langle\hat{J}_{X}\rangle$. This
is a generic property of planar squeezed minimum variance states,
which always have a spin vector with a finite mean amplitude in the
plane of minimum variance. 

The value of the spin variance in each direction in the plane still
needs to be calculated, in order to define the properties of the planar
squeezed state. We assume a symmetric amplitude distribution with
$R_{m}=R_{-m}$, so that $\langle\hat{J}_{Z}\rangle=0$. The mean
values are then given as in equation (\ref{eq:meanvalue}), by:

\begin{eqnarray}
\langle\hat{J}_{X}\rangle & = & \frac{1}{n}\sum_{M=-\tilde{J}}^{M=\tilde{J}}R_{M_{+}}R_{M_{-}}\sqrt{\tilde{J}^{2}-M^{2}}\,,\nonumber \\
\langle\hat{J}_{Y}\rangle & = & \langle\hat{J}_{Z}\rangle=0\,.
\end{eqnarray}

Introducing $J_{+}=J+1$ , $m_{+}=m+1$, we find the squared spin
projections and correlations are:

\begin{eqnarray}
\langle\hat{J}_{X}^{2} & \rangle= & \frac{1}{2n}\{\sum_{m=-J}^{J}\left[J^{2}-m^{2}+J\right]R_{m}^{2}\\
 &  & +\sum_{m=-J}^{J-2}\sqrt{\left[J_{+}^{2}-m_{+}^{2}\right]\left[J^{2}-m_{+}^{2}\right]}R_{m}R_{m+2}\}\,,\nonumber 
\end{eqnarray}

\begin{eqnarray}
\langle\hat{J}_{Y}^{2} & \rangle= & \frac{1}{2n}\{\sum_{m=-J}^{J}\left[J^{2}-m^{2}+J\right]R_{m}^{2}\\
 &  & -\sum_{m=-J}^{J-2}\sqrt{\left[J_{+}^{2}-m_{+}^{2}\right]\left[J^{2}-m_{+}^{2}\right]}R_{m}R_{m+2}\}\,,\nonumber 
\end{eqnarray}

\begin{equation}
\langle\hat{J}_{Z}^{2}\rangle=\frac{1}{n}\sum_{m=-J}^{J}m^{2}R_{m}^{2}\,,
\end{equation}

\begin{equation}
\langle\hat{J}_{X}\hat{J}_{Y}\rangle=\langle\hat{J}_{X}\hat{J}_{Z}\rangle=\langle\hat{J}_{Z}\hat{J}_{Y}\rangle=0\,.
\end{equation}

We now treat the asymptotic, large $J$ limit using an integral approximation
as previously. For the mean value term, on replacing summations by
integrals, in the limit of $J\rightarrow\infty$ and defining $x=M/\tilde{J}$,
$r(x)=R_{M}\sqrt{\tilde{J}}$ and $\bar{J}_{X}=\langle J_{X}\rangle$
we obtain: 
\begin{eqnarray}
\bar{J}_{X} & = & \int_{-1}^{1}\left(r^{2}(x)-\frac{\left(r'(x)\right)^{2}}{2\tilde{J}^{2}}+\frac{xr(x)r'(x)}{4J^{2}(x)}\right)J(x)dx\,,\nonumber \\
 &  & \,
\end{eqnarray}
where $J(x)\equiv\tilde{J}\sqrt{1-x^{2}}$. We know that $\sigma=\left(2J\right)^{-2/3}$
to get the minimum variance. To leading order, one finds that:
\begin{equation}
\bar{J}_{X}=\tilde{J}\int_{-1}^{1}r^{2}(x)\left(1-\frac{x^{2}}{8\sigma^{2}\tilde{J}^{2}}\right)J(x)dx\,,
\end{equation}
Next, we expand the variational integral, retaining leading terms
only, giving
\begin{equation}
\bar{J}_{X}\approx\tilde{J}\left[1-\frac{\sigma}{2}-\frac{3\sigma^{2}}{8}-\frac{1}{8\sigma\widetilde{J}^{2}}\right]\sim J\left[1-\frac{1}{2\left(2J\right)^{2/3}}\right]\label{eq:jx}
\end{equation}
Similarly, on evaluating the square of the z-projection, we find that:
\begin{equation}
\langle\hat{J}_{Z}^{2}\rangle=\tilde{J}^{2}\sigma+\frac{1}{4}\sim\left(J^{2}/2\right)^{2/3}\,.
\end{equation}

The sum of planar variances is simple to calculate. Following the
techniques given previously, we find that:
\begin{equation}
\langle\hat{J}_{X}^{2}+\hat{J}_{Y}^{2}\rangle=\tilde{J}^{2}(1-\sigma)-\frac{1}{2}\,.
\end{equation}
This means that we carry out a check on the total spin, which is given
by the expected result of:

\begin{eqnarray}
\langle\hat{J}_{X}^{2}+\hat{J}_{Y}^{2}+\hat{J}_{Z}^{2} & \rangle= & \tilde{J}^{2}-\frac{1}{4}\nonumber \\
 & = & J(J+1)\,.
\end{eqnarray}

However, the individual planar variances are more complex. Introducing
scaled variables of $x=(m+1)/J=M'J$ , one must use a Taylor expansion
so that: 
\begin{align}
r(x\pm\frac{1}{J}) & =r(x)\pm\frac{1}{J}r'(x)+\frac{1}{2J^{2}}r''(x)+O(\frac{1}{\tilde{J}^{3}})\,\nonumber \\
 & =r(x)\mp\frac{x}{2J\sigma}r(x)-\frac{r(x)}{4J^{2}\sigma}+\frac{x^{2}r(x)}{8J^{2}\sigma^{2}}+O(\frac{1}{\tilde{J}^{3}})\,.
\end{align}
After carrying out the integrals, we find that the squared projections
in the plane are: 
\begin{eqnarray}
\langle\hat{J}_{X}^{2}\rangle & = & J^{2}\left[1-\sigma-\frac{1}{4J^{2}\sigma}+\frac{1}{J}+\frac{1}{4J^{2}}\right]\nonumber \\
\langle\hat{J}_{Y}^{2}\rangle & = & \frac{1}{4}\left[\frac{1}{\sigma}-1\right]\,.
\end{eqnarray}
on subtracting the mean values squared, this gives the result that
to leading order:

\begin{eqnarray}
\Delta^{2}J_{X} & \sim & \frac{\left(2J\right)^{2/3}}{8}\,,\nonumber \\
\Delta^{2}J_{Y} & \sim & \frac{\left(2J\right)^{2/3}}{4}\,,\nonumber \\
\Delta^{2}J_{Z} & \sim & \left(J^{2}/2\right)^{2/3}\,,\label{eq:VarJx}
\end{eqnarray}
and hence the Heisenberg uncertainty principle in the $Z-Y$ plane
is obeyed, since:
\begin{equation}
\Delta J_{Z}\Delta J_{Y}\sim\frac{J}{2}\ge|\langle\hat{J}_{Z}\rangle|/2\,.\label{eq:varJy}
\end{equation}

On calculating the mean and variance of each spin component, we find
that that the plane of squeezing always includes the mean spin direction
$\langle\hat{\mathbf{J}}\rangle$. Choosing axes so that this projection
is in the $X$ direction, with planar squeezing in the $X-Y$ plane,
we calculate that: $\langle\hat{J}_{X}\rangle\sim J-\frac{1}{2}\left(J/4\right)^{1/3}\,,$
$\Delta^{2}J_{X}\sim\frac{1}{8}\left(2J\right)^{2/3}\,,$ $\Delta^{2}J_{Y}\sim\frac{1}{4}\left(2J\right)^{2/3}\,,$
$\Delta^{2}J_{Z}\sim\left(J^{2}/2\right)^{2/3}\,$ . Asymptotically
this is a Heisenberg limited or `intelligent' \cite{intangmom} state
in the $Z-Y$ plane since: 
\begin{equation}
\Delta J_{Y}\Delta J_{Z}\approx\frac{1}{2}\left|\langle\hat{J}_{X}\rangle\right|.
\end{equation}
\[
\]

In summary, the important features of the optimum planar squeezed
state are:
\begin{itemize}
\item A large spin expectation \emph{parallel }to the plane of squeezing
\item Maximum variance reduction in the direction of the mean spin vector
\item A smaller variance reduction in the plane of squeezing orthogonal
to the mean spin
\item Saturation of the $C_{J}$ uncertainty relation 
\item A complementary variance increase in the third dimension
\item Heisenberg limited asymptotic uncertainties \emph{perpendicular} to
the mean spin vector
\end{itemize}

\section{Applications of Planar Quantum Squeezing}

\subsection{Generation of PQS in a BEC}

\begin{figure}
\begin{centering}
\includegraphics[width=1\columnwidth]{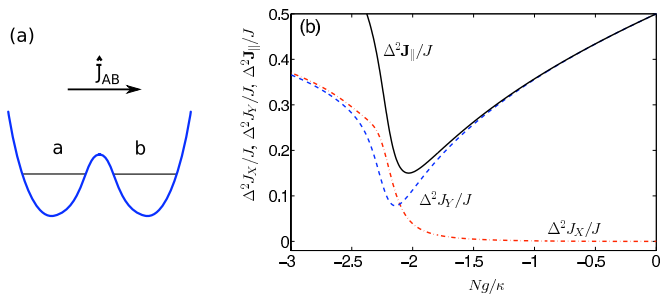}
\par\end{centering}

\caption{(Color online) (a) Schematic diagram of double-well Bose-Einstein
condensate\textbf{.} Shows a schematic double potential well, with
corresponding pseudo-spin operator. (b) Spin variances for the ground
state of a double-well BEC. Here $N=100$ atoms, or $J=50$. The solid
line is the total variance, the dashed line is the Y-axis spin projection
variance, the dot-dashed line is the $X$- axis variance in the mean
spin direction. \label{fig:Schematic-diagram-of}}
\end{figure}

We first wish to consider how to generate these PQS states. While
there are many possible strategies, the simplest is to find a physical
system whose Hamiltonian equals the variance sum. The ground state
of such a Hamiltonian will minimize the variance, hence creating a
perfect planar squeezed state. This can be readily achieved in a two-mode
Bose-Einstein condensate, which has been experimentally demonstrated
to generate spin-squeezing \cite{oberthaler,oberthaler2010}. In the
limit of tight confinement and small numbers of atoms, this type of
system can be treated using a simple coupled mode effective Hamiltonian,
where $\kappa$ is the inter-well tunneling rate between wells, and
$g$ is the intra-well s-wave interaction between the atoms. The system
is depicted schematically in Fig. \ref{fig:Schematic-diagram-of}.
We note at that the total particle number is conserved, with eigenvalue
$N=2J$, where $J$ is the equivalent effective spin quantum number. 

Following standard techniques \cite{He-PRL,Hines,XieHai}, the two-well
BEC Hamiltonian can be written in spin language, ignoring conserved
terms proportional to $\hat{N}$ or $\hat{N}^{2}$. We start with
a two-mode Hamiltonian in the standard form:

\begin{equation}
\hat{H}/\hbar=\kappa\left[\hat{a}^{\dagger}\hat{b}+\hat{b}^{\dagger}\hat{a}\right]+\frac{g}{2}\left[\hat{a}^{\dagger}\hat{a}^{\dagger}\hat{a}\hat{a}+\hat{b}^{\dagger}\hat{b}^{\dagger}\hat{b}\hat{b}\right]\,.\label{eq:Hamiltonian}
\end{equation}

Inter-well spin operators have already been measured in this environment,
and are defined as:
\begin{eqnarray}
\mathbf{\hat{J}}_{\parallel} & = & \left(\frac{1}{2}\left[\hat{a}^{\dagger}\hat{b}+\hat{a}\hat{b}^{\dagger}\right],\frac{1}{2i}\left[\hat{a}^{\dagger}\hat{b}-\hat{a}\hat{b}^{\dagger}\right]\right)\,,\nonumber \\
\hat{J}_{\perp} & = & \left(\hat{a}^{\dagger}\hat{a}-\hat{b}^{\dagger}\hat{b}\right)/2\,,\label{eq:Spinmapping}
\end{eqnarray}
 It is convenient to introduce a symmetry breaking vector $\mathbf{J_{0}}=\left(\kappa/g,0\right)$,
which causes tunneling, to give:
\begin{equation}
\hat{H}/\hbar=-g\left|\mathbf{\hat{J}}_{\parallel}-\mathbf{J_{0}}\right|^{2}\,.\label{eq:Hamiltonian-1}
\end{equation}

It is clear from this form of the Hamiltonian that with an attractive
coupling so that $g<0$, the ground state will \emph{exactly} minimize
the planar variance, provided the tunneling rate $\kappa$ is adjusted
so that $\left\langle \mathbf{\hat{J}}\right\rangle =\mathbf{J_{0}}\,.$
Since we have already calculated the optimum mean spin vector, we
therefore expect that a planar squeezed state will occur as the ground
state of the Hamiltonian for an attractive coupling ($g<0$), with
$\kappa/\left|g\right|=\left|\left\langle \hat{J}_{X}\right\rangle \right|=J-\frac{1}{2}\left(J/4\right)^{1/3}$.
The eigenstates of this Hamiltonian can be readily calculated numerically
from its matrix form. It is known to have a macroscopic inter-well
spatial entanglement \cite{Vedral,carr macrosup,He-PRL}, which is
maximized at a critical attractive coupling value \cite{He-PRL,Hines,XieHai}. 

A graph of the variances against coupling is shown in Fig. \ref{fig:Schematic-diagram-of}b,
showing the expected reduction in both variances at a critical value
of the coupling. For $N=100$, the total planar spin variance reaches
its minimum value with a coupling value of $Ng/\kappa=-2.034=N/\left\langle \hat{J}_{X}\right\rangle $,
as expected. At the minimum variance, we find numerically that: 
\begin{equation}
\Delta^{2}\mathbf{J}_{\parallel}=0.1501\, J=C_{50}\,.
\end{equation}
This is in excellent agreement with equation (\ref{eq:Cj}), which
gives $C_{50}^{(a)}=0.1499\, J$. Similar good agreement is obtained
for the calculated values of $\left\langle \hat{\mathbf{J}}_{X}\right\rangle ,$
$\Delta^{2}\mathbf{J}$, compared with the asymptotic equations (\ref{eq:jx}),
(\ref{eq:VarJx}), and (\ref{eq:varJy}) - apart from corrections
of order $1/J^{2}$.

In summary, the ground state of a two-well BEC is not only a planar
squeezed state: at critical coupling it gives the \emph{exact} solution
to the minimum variance. However, while spin-squeezing has been observed
experimentally \cite{oberthaler,oberthaler2010,atom-chip}, indicating
that detection at the quantum shot-noise level is technically feasible,
existing experiments used a BEC with repulsive interactions. To obtain
a PQS state would require an experiment using attractive interactions,
as found in $^{39}K$, for example \cite{Modugno}. We finally note
that Einstein-Podolsky-Rosen entanglement and macroscopic superpositions
for BEC states have been topics of recent interest \cite{He-PRL,ferriscaval,brandjoch,isreprbecprl}.
It is intriguing that the simplest physical route towards generating
planar-squeezed states also displays other important physical properties,
including macroscopic entanglement \cite{He-PRL}. In this case we
emphasize that the entanglement is found between the two underlying
boson modes which are used to construct the composite spin operators.

\subsection{Applications to Interferometry}

Due to their interactions with magnetic and gravitational fields,
cold atomic sensors have been useful for ultra-sensitive magnetometers
\cite{SpinorMagnetometer} and gravimeters \cite{ChuGravimeter,Close}.
In this section, we analyze the possible applications of planar spin
squeezing in atom interferometry \cite{PritchardRMP}. In order to
improve the performance of this type of sensor it is important to
combine both relatively large atom numbers\cite{Sidorov} and quantum
noise reduction\cite{oberthaler2010,atom-chip}. Consider the effect
of the spin mapping to a pair of boson modes which are subsequently
passed through a beam splitter (for external degrees of freedom) or
microwave rotation (in the case of internal degrees of freedom) \cite{spinsqkiti,spinsqwine}.
We use the mapping defined in the previous section in equation (\ref{eq:Spinmapping}),
with a planar quantum squeezed input.

\begin{figure}
\begin{centering}
\includegraphics[width=0.8\columnwidth]{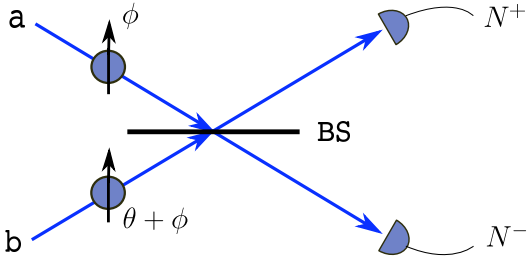}
\par\end{centering}

\caption{(Color online) Schematic diagram of ideal interferometric measurements.
Phase measurements can be reduced in the simplest case to an output
measurement of number differences from a beam-splitter, where $\phi$
is the unknown phase-shift and $\theta$ is a known reference phase-shift.\label{fig:Interferometric-measurements}}
\end{figure}

Next, we consider the effects of an idealized phase-measurement. This
is shown in schematic form in Fig \ref{fig:Interferometric-measurements}.
Two bosonic modes $\hat{a}$, $\hat{b}$ prepared in an entangled
PQS state, are passed through phase-shifters and a four-port $50:50$
beam-splitter. Here $\phi$ is an unknown phase-shift to be measured,
while $\theta$ is a reference phase-shift. The number difference
between the output ports, $N^{+}-N^{-}$ is measured, and gives information
about the unknown phase. 

After the phase shifters, but before the beam splitter, the collective
spin components in the $X-Y$ plane are:
\begin{eqnarray}
\hat{J}_{X}(\phi) & = & \hat{J}_{X}cos\left(\phi-\theta\right)+\hat{J}_{Y}sin\left(\phi-\theta\right)\,,\nonumber \\
\hat{J}_{Y}(\phi) & = & \hat{J}_{Y}cos\left(\phi-\theta\right)-\hat{J}_{X}sin\left(\phi-\theta\right)\,.
\end{eqnarray}
After the beam splitter, we define two output operators as
\begin{eqnarray}
\hat{c}_{\pm} & = & \left[\hat{a}e^{i\phi}\pm\hat{b}e^{i\theta}\right]/\sqrt{2}\,.
\end{eqnarray}
Calculating the phase-sensitive output number difference, $\hat{\mathcal{N}}=\left[\hat{N}^{+}-\hat{N}^{-}\right]$,
and its derivative in terms of the measured phase, we find that the
measured phase uncertainty is:
\begin{equation}
\Delta\phi=\frac{\sqrt{\Delta^{2}\mathcal{N}(\phi)}}{\left|\partial\bar{\mathcal{N}}/\partial\phi\right|}
\end{equation}

The phase noise in a single measurement in terms of the initial spin
variances, in a state where $\langle J_{X}J_{Y}\rangle=0$, is therefore:
\[
\Delta\phi=\frac{\sqrt{\Delta^{2}J_{X}cos^{2}\left(\phi-\theta\right)+\Delta^{2}J_{Y}sin^{2}\left(\phi-\theta\right)}}{\left|\bar{J}_{Y}cos\left(\phi-\theta\right)-\bar{J}_{X}sin\left(\phi-\theta\right)\right|}\,.
\]
Clearly, in any phase measurement one must avoid insensitive regimes
of the interferometer near the fringe peaks where $\bar{J}_{Y}(\phi)=\partial\bar{\mathcal{N}}/\partial\phi\sim0$.
We see that the quantum noise in each measurement is bounded above
since:
\begin{equation}
\Delta^{2}J_{X}cos^{2}\left(\phi-\theta\right)+\Delta^{2}J_{Y}sin^{2}\left(\phi-\theta\right)\le\Delta^{2}\mathbf{J}_{\parallel}\,.
\end{equation}
This shows clearly that the planar variance is an upper bound to the
total quantum noise on the interferometer number difference output.
Squeezing this upper bound below the standard quantum limit is vital
under conditions where the the measured phase is completely unknown.
Planar quantum squeezed states therefore can provide useful noise
reductions over a large range of unknown phases. In the best case,
one can achieve $\Delta^{2}\mathbf{J}_{\parallel}\sim C_{J}\sim J^{2/3}$,
which leads to a phase uncertainty of: 
\begin{equation}
\Delta\phi\propto J^{-2/3}.
\end{equation}

This has the utility of allowing much lower atom numbers and therefore
lower atomic density in atom interferometry at a given phase sensitivity.
Since the atom density is limited by two-body and three-body losses,
this is a very significant practical advantage. To give an example,
a PQS interferometer with $10^{6}$ atoms has a phase sensitivity
of order $10^{-4}$ in a single phase measurement. To achieve this
result with conventional coherent interferometry would require $10^{8}$
atoms, which implies $100$ times greater density for the same geometry.

\section{Planar squeezing entanglement criteria}

In earlier works, uncertainty relations have been used extensively
for the derivation of criteria to detect entanglement, the EPR paradox
and mesoscopic superpositions \cite{hoftoth,spinsq,toth2009,eprbohmparadox,hw-1-1,wiseman2007,expphotonweinfurt,tothjosaBdickefour,mdr duan uncert,reidrmp,sumuncerduan,spinprodg,Kdechoum,soresinsqent,toth,magsusentspin,spinsqkorb,pappkimblevar,Anders_entanglement,cavalreiduncer,cavalreiduncer2006}.
However, neither the Heisenberg \cite{spinprodg} nor the related
Sorenson-Molmer \cite{soresinsqent,Anders_entanglement} inequalities
can be used to detect the entanglement of a very important subclass
of states that have $\langle\mathbf{J}\rangle=0$. This includes the
maximally entangled states which are eigenstates of $J_{Z}$ having
$\langle\hat{J}_{Z}\rangle=0$. In these cases, that give rise to
the classic violations of local realism studied originally by Bell
\cite{Bell,key-6}, the variance of the remaining spins are still
constrained, because no simultaneous eigenstates exist. 

It is this situation our more general two-component uncertainty relation
involving both variances is extremely useful \cite{hoftoth}. This
two-component criterion can also detect the entanglement of the Bell-type
maximally entangled states, only requiring the measurement of two
orthogonal spin components \cite{hoftoth}, as well as determining
the possible existence of multi-particle genuine entanglement. A further
application of the planar uncertainty relation equation (\ref{eq:twospinsumcj})
is that it provides a means of witnessing multipartite entanglement
between macroscopic spins. 

Hofmann and Takeuchi \cite{hoftoth} have proved that any uncertainty
relation of the type $\Delta^{2}\mathbf{A}\geq U_{A}$, where the
system is labelled $A$ and $\mathbf{A}$ is a vector of observables
for that system, can be used to define a criterion for entanglement.
Here the limit $U$ is generically defined as the absolute minimum
of the uncertainty sum for any quantum state. The planar uncertainty
equation (\ref{eq:twospinsumcj}) is of this form. If systems $A$
and $B$ are separable, then it is always true that 
\begin{equation}
\Delta^{2}\left(\mathbf{A}+\mathbf{B}\right)\geq U_{A}+U_{B}\,.
\end{equation}

The violation of this uncertainty bound is then a proof of entanglement.
This result may be generalized to multipartite systems consisting
of $N$ distinct locations, as shown by Toth \cite{spinsq,toth2009,toth,tothmit}.
Consider $N$ sites of spin $J$ particles. Assuming absolute separability,
we write the total density matrix $\hat{\rho}$ as a probabilistic
sum of product density matrices $\hat{\rho}_{R}^{k}$ at site $k$,
occurring with probability $P_{R}$: 
\begin{equation}
\hat{\rho}=\sum_{R}P_{R}\hat{\rho}_{R}^{1}\hat{\rho}_{R}^{2}...\hat{\rho}_{R}^{N}\,.\label{eq:sepN}
\end{equation}
Defining collective spin operators as
\begin{equation}
\hat{J}_{i}^{total}=\sum_{k=1}^{N}c_{k,i}\hat{J}_{i}^{k},
\end{equation}
where $c_{k,i}=\pm1,$ the expression for the sum of the variances
is
\begin{equation}
\Delta^{2}\mathbf{J}_{\parallel}^{total}\geq\sum_{R}P_{R}\sum_{k=1}^{N}(\Delta^{2}\mathbf{J}_{\parallel}^{k})_{R}\,.
\end{equation}
\[
\]
Using the two-component uncertainty equation (\ref{eq:twospinsumcj}),
this leads to the separability condition for $S_{2}$, the sum of
the two-component relative variances: 
\begin{equation}
S_{2}=\Delta^{2}\mathbf{J}_{\parallel}^{total}\geq NC_{J}\,,\label{eq:hofmann-1-1}
\end{equation}
If violated, this will imply an entanglement between some of the sites.
When three variances are measurable, there are similar conditions
known based on three-variance uncertainties.

These relations are useful for identifying the entanglement of maximally
entangled states where $\langle\hat{\mathbf{J}}\rangle=0$. For example,
the Bell singlet-state $J=1/2$ for which the spins are anti-correlated
($J^{total}=0$) gives the total uncertainty as zero (i.e., $S_{2}=0$)
provided $c_{k}=d_{k}=e_{k}=1$. The many body singlet state in which
pairs of particles are in singlet states gives zero total uncertainties
for larger $N$ \cite{tothmit}. Here we consider the maximally entangled
states for fixed $J$ of form (here $|J,\, m\rangle$ are the eigenstates
of $J_{Z}$)
\begin{equation}
|J,N\rangle_{M}=\frac{1}{\sqrt{2J+1}}\sum_{m=-J}^{J}|J,\, m\rangle^{\otimes N}\,,
\end{equation}
where spins are correlated. For $N=2$, the sum of the two variances
$\Delta\left(J_{X}^{A}-J_{X}^{B}\right)^{2}+\Delta\left(J_{Y}^{A}+J_{Y}^{B}\right)^{2}$
is always zero, for any $J$. Singlet states of total spin zero for
$d$-level systems have been presented in \cite{cabsupersinglet}.
We denote the singlet state (which has zero total spin) obtained from
$N$ particles of spin $J$ as $|J,N\rangle_{S}$.

The two variance sum $S_{2}$ is zero for the anti-correlated singlet
state. To examine the sensitivity of the two variance criterion to
noise, we reduce entanglement by considering the mixed state of the
type considered by Werner \cite{Werner}:
\begin{equation}
\hat{\rho}=p_{n}\hat{I}_{J}+\left(1-p_{n}\right)|J,N\rangle_{S}\,_{S}\langle J,N|\,,
\end{equation}
where $p_{n}$ gives the relative contribution of the white noise
term represented by $\hat{I}_{J}=\left[\hat{I}/\left(2J+1\right)\right]^{\otimes N}$,
which is a rotationally symmetric, uncorrelated state proportional
to the identity matrix at each site. Since the singlet state is perfectly
correlated, and the Werner state is completely isotropic, one can
show that the uncertainties for the multipartite case are entirely
due to the white noise terms. From equation (\ref{eq:hofmann-1-1}),
which requires only two measurement settings, the condition for detecting
entanglement is given by:
\begin{equation}
S_{2}=\frac{2N}{3}J(J+1)p_{n}<NC_{J}\,.
\end{equation}
which gives the bound of noise for detecting entanglement of:
\begin{equation}
p_{n}<\frac{3C_{J}}{2J\left(J+1\right)}\,.
\end{equation}
If all three spin measurements are available, a three spin measurement
is even more sensitive to entanglement for this isotropic case. However,
in many cases all three components are not measurable, or may not
all have strong correlations. 

\begin{figure}
\begin{raggedright}
\includegraphics[width=0.8\columnwidth]{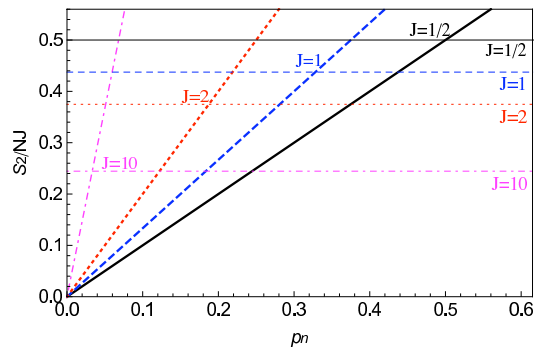}
\par\end{raggedright}

\caption{(Color online) Plot of the multipartite entanglement criteria. The
sloping, bold lines give the normalized sum of the two spin variances
($S_{2}/NJ$) for the $N$ particle singlet state $|J,N\rangle_{S}$
mixed with the maximally noisy state, as a function of noise probability
$p_{n}$. Entanglement is confirmed when the bold line for each $J$
falls below the corresponding horizontal line giving $C_{J}/J$. From
the bottom, the lines correspond to $J=1/2,1,2,10$.\label{fig:cj-2}}
\end{figure}

Figure \ref{fig:cj-2} illustrates the purity required in order to
demonstrate entanglement between spin $J$ particles (sites) using
the two variance equation (\ref{eq:hofmann-1-1}). Entanglement is
detected when the normalized two variance sum $S_{2}$ is below the
line indicating the value of $C_{J}/J$, i.e., when $S_{2}/NJ<C_{J}/J$.

\section{Outlook}

In conclusion, we have derived an uncertainty relation for the planar
sum of the variances in two orthogonal spin directions for systems
of fixed total spin. The lower bound varies asymptotically as $J^{2/3}$.
We have shown that this planar local uncertainty relation can be readily
saturated with macroscopic planar squeezed states at large spin. In
addition, a practical technique for generating these states is proposed,
employing the ground state of a two-well Bose condensate with attractive
interactions. 

Such planar squeezed states have the feature that they are able to
minimize phase measurement noise over a wide range of unknown phase
angles, in an interferometric measurement. Since these states are
readily obtainable as ground states of physically relevant Hamiltonians
in a two-mode Bose-Einstein condensate, it appears feasible to generate
and demonstrate these features in laboratory measurements, either
using ground state preparation in an attractive BEC \cite{He-PRL},
or using dynamical techniques with repulsive interactions.

Criteria for detecting entanglement between multiple spin $J$ systems
using only two component measurements can be derived from this. The
two spin component entanglement criterion is likely to have important
applications where noise or measurement is asymmetric, so that all
three components cannot be measured. There are other generalizations
possible if one constrains $\langle\hat{J}_{Z}\rangle$ to have a
finite value, giving an inequality involving $C_{J}\left(\bar{J}_{Z}\right)$;
these will be treated elsewhere. We expect similar relations to occur
for other continuous groups.
\begin{acknowledgments}
We wish to thank the Humboldt Foundation, Heidelberg University, and
the Australian Research Council for funding via AQUAO COE and Discovery
grants, as well as useful discussions with Markus Oberthaler, Philip
Treutlein and Andrei Sidorov. \end{acknowledgments}

\end{document}